\documentclass[]{interact}

\usepackage[T1]{fontenc}
\usepackage{epstopdf}
\usepackage{graphicx}
\usepackage[caption=false]{subfig}
\usepackage[numbers,sort&compress]{natbib}
\usepackage{placeins}
\usepackage{url}
\usepackage{xcolor}
\definecolor{linkblue}{RGB}{0,90,160}
\usepackage[colorlinks=true,linkcolor=linkblue,citecolor=linkblue,urlcolor=linkblue]{hyperref}
\bibpunct{[}{]}{,}{n}{,}{,}

\makeatletter
\def\NAT@def@citea{\def\@citea{\NAT@separator}}
\makeatother

\newcommand{\hmin}{h_{\min}}
\newcommand{\ttc}{(t_0-t)}
\newcommand{\Ca}{\mathrm{Ca}}
\newcommand{\Oh}{\mathrm{Oh}}
\newcommand{\We}{\mathrm{We}}

\newcommand{\De}{\mathrm{De}}

\newcommand{\ellnu}{\ell_\nu}
\newcommand{\thetaa}{\theta_{\mathrm{ap}}}
\newcommand{\thetae}{\theta_{\mathrm{e}}}

\begin{document}

\articletype{Review Article}

\title{Singularities in Soft Matter Systems}

\author{
\name{Vatsal Sanjay\thanks{CONTACT Vatsal Sanjay. Email: vatsal.sanjay@comphy-lab.org}}
\affil{CoMPhy Lab, Department of Physics, Durham University, Science Laboratories, South Road, Durham DH1 3LE, United Kingdom}
}

\maketitle

\begin{abstract}
When a liquid thread pinches off, its neck thins as it separates into two unconnected regions.
Using continuum mechanics, we can predict that the neck reaches zero radius in finite
time while its curvature grows without bound. Together, the vanishing neck and diverging
curvature form a finite-time singularity. However, a real fluid
does not realise these mathematical limits as molecular or material physics takes over once
the neck becomes sufficiently small.
Similar singularities arise throughout soft matter whenever a smooth continuum description is used at vanishing length scales.
This review asks what the shrinking region forgets, what it retains, and which material length, time, or stress cuts off the apparent divergence.
The dynamics near a singularity often become self-similar, with profiles at different times
collapsing onto one shape when rescaled by the shrinking local length.
Sometimes that collapse is universal enough that the surrounding geometry and forcing no longer determine the local dynamics.
Nonetheless, the measured output could still depend on how the
shrinking region is fed by the surrounding flow and on the small-scale physics that finally
replaces the ideal divergence.
Complex fluids and active matter change the same local balance by bringing
their own timescales into the shrinking region.
Beyond interfaces, the same logic applies when the localised object is a
stress concentration or a defect in geometry or order rather than a moving
surface.
Singularities matter because they show where continuum theory stops being
the relevant description and how the small-scale cutoff sets the outputs that
count in printing, coating, aerosols, and stretchable solids.
\end{abstract}


\section{Soft matter and singularities}

\begin{figure}
\centering
\includegraphics[width=\textwidth]{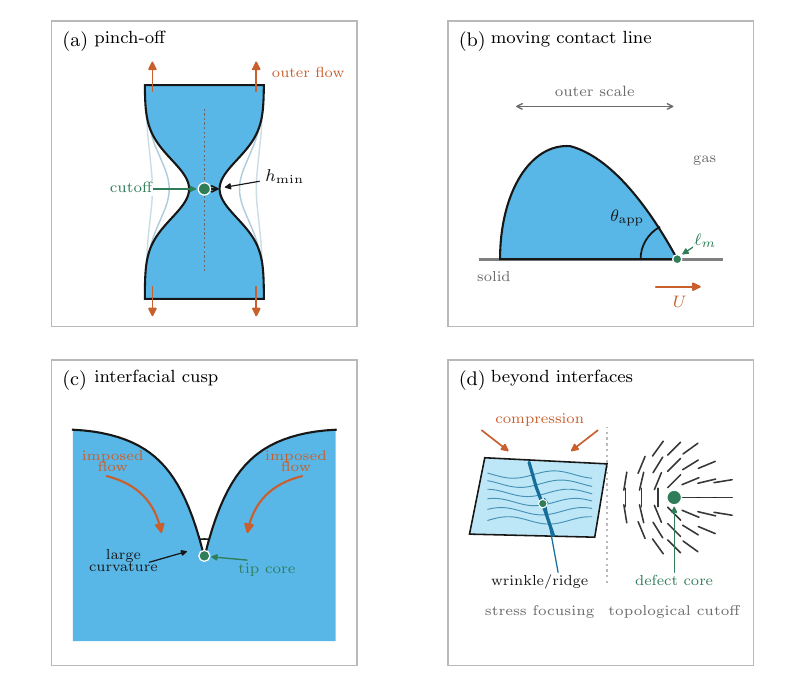}
\caption{A visual dictionary for singularities in soft matter. The four panels are
schematics of different singular events: (a)~Drop pinch-off. An outer flow stretches a liquid thread whose neck
radius $\hmin$ is driven towards zero, drawn with faint self-similar profiles around the neck
and a cutoff at the core. (b)~A moving contact line. A drop advances at speed $U$ over a
solid, and the apparent contact angle $\thetaa$ is set by matching the outer scale down to a
microscopic cutoff $\ell_m$ at the three-phase line. (c)~An interfacial cusp. An imposed flow
pulls a free surface into a sharp tip of large curvature, regularised at a small tip core.
(d)~Beyond interfaces. Compression focuses stress in a thin elastic sheet into a wrinkle or ridge, while an ordered soft material forms a defect with a finite core; the sheet thickness and core size
set the respective cutoffs.}
\label{fig:visual-dictionary}
\end{figure}

Soft matter is often introduced through deformability.
For example, modest external forces and thermal fluctuations can dictate how droplets,
bubbles, gels, or other soft materials deform.
This contrasts with hard matter, where deformation is usually resisted by strong interatomic bonding
and comparatively rigid microscopic structure.
Yet softness does not imply smoothness as these deformations can localise.
Consequently, a smooth continuum field can focus into a vanishing length or time scale and develop an
unbounded derivative. These limiting events are what we call singularities.
Here, we stress that we use the term operationally. It is not a claim that nature produces a
literal infinity, but an ideal prediction made by a continuum theory when it runs out of
scale.

Such limits are ubiquitous. At a dripping faucet, just before a drop falls, it hangs from
a capillary neck whose radius is driven towards zero within the macroscopic description
(figure~\ref{fig:visual-dictionary}a). Coalescence begins from the other side of the same
topological change, when two separated drops first communicate through a bridge set by
their first contact. In both cases, the ideal
continuum description sends the interfacial curvature to infinity.
Here, topology is a proxy for connectivity. Pinch-off changes one connected liquid body into two, whereas coalescence reverses the change.
In ordered soft materials, topology can instead force an orientation or phase field to become
undefined at a defect; its finite core regularises the ideal singularity.
A moving solid--liquid--gas contact line, such as the rim of a raindrop sliding on a
window pane, exposes a different divergence where the macroscopic viscous stress becomes
infinite (figure~\ref{fig:visual-dictionary}b). A viscous free surface can also make a
singular geometry visible by sharpening into a cusp, where the curvature diverges at the
tip (figure~\ref{fig:visual-dictionary}c). Beyond liquids, localisation also appears in a
stretched elastic sheet that wrinkles to relax compressive stress by bending out of plane
(figure~\ref{fig:visual-dictionary}d). Indeed, away from the singular region, a continuum equation can be an excellent description.
The examples above illustrate how its failure can be local.
While singularities can naively be considered a flaw in the calculations,
they often tell us where the dominant physics has become too approximate, and
therefore what ingredient the theory has missed.
Slip, gas compressibility, thermal noise, intermolecular forces, surface chemistry, finite extensibility, or elasticity may
eventually regularise the predicted divergence by introducing a finite small-scale cutoff.

Many physicists are first introduced to a singularity through black holes. In general
relativity, the spacetime curvature becomes unbounded. Here, the infinity matters because
it marks the failure scale of a classical theory. Eggers and
Fontelos \citep{EggersFontelos2015Singularities} use this emblem as an entry point to a
wider view where shocks, caustics, cracks, vortices, contact lines, and pinching drops are
also singular structures of continuum partial differential equations. In soft matter the
diverging quantity often appears directly as interface curvature, viscous stress, elastic strain,
or a gradient of an order parameter. Eggers and Fontelos call this local structure the
``fingerprint'' of a nonlinear partial differential equation
\citep[p.~4]{EggersFontelos2015Singularities}. As a length scale becomes small, many
remote details fall away and the dominant balance is exposed.

For this review, we primarily use examples from interfacial flows as they make
the abstract mathematical treatment of singularity visible in familiar experiments
\citep[see figure~\ref{fig:visual-dictionary} and][]{Eggers1997Nonlinear,EggersVillermaux2008PhysicsLiquidJets,EggersFontelos2015Singularities}.
For example, a thinning capillary bridge makes the shrinking length scale visible as
its neck approaches the endpoint predicted by the ideal continuum description. Contact
lines and cusps expose the same breakdown through diverging stress and curvature,
respectively. In each case, we consider both the approach to the singular limit and the
physical mechanism that regularises it. The same framework then extends to elastic
sheets, defects, and soft solids.

This review is arranged as follows.
Section~\ref{sec:toolkit} introduces the toolkit used throughout the paper, from the
shrinking length scale and dominant balance to universality, material memory, and
regularisation. Sections~\ref{sec:drop-pinch}--\ref{sec:jetting} examine drop
pinch-off, bubble pinch-off, coalescence, contact lines, cusps, thin films, and jets born
from singular collapse. Section~\ref{sec:complex} considers how polymers, elasticity,
particles, or surfactants allow the fluid to retain material memory. Section~\ref{sec:beyond} extends
the framework to soft solids and ordered media, and section~\ref{sec:end} brings these
examples together in a synthesis figure and considers their applications.

\section{A toolkit for singular thinking}
\label{sec:toolkit}

\begin{figure}
\centering
\includegraphics[width=\textwidth]{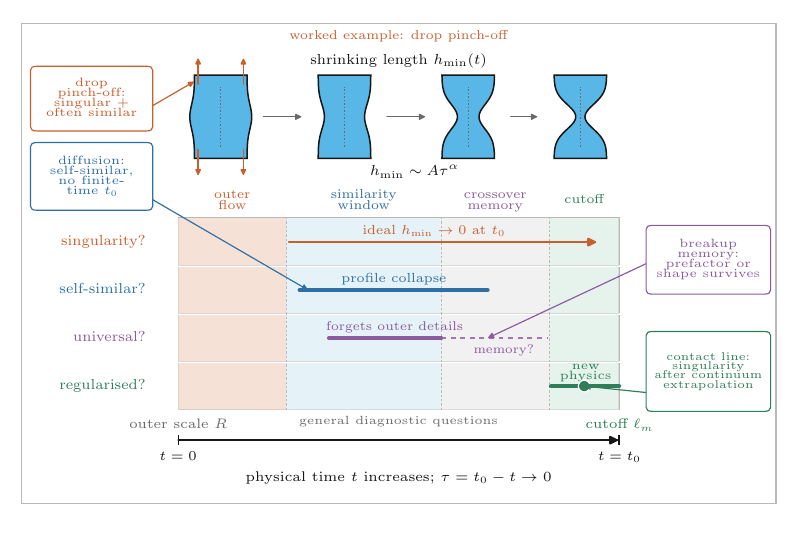}
\caption{The conceptual toolkit, drawn as a reference card for the sections that follow. A
candidate singularity is read along a route in physical time~$t$ (horizontal axis), from the
outer scale~$R$ at $t=0$ towards the cutoff~$\ell_m$ as the time-to-singularity
$\tau\equiv t_0-t\to0$. The thinning-neck sequence above is a worked example of drop
pinch-off, with $\hmin\sim A\tau^\alpha$; the four lanes below are general diagnostic
questions, while the side callouts show how the same toolkit applies to this and other processes.
The \emph{singularity?} lane asks which length vanishes and where
the smooth description fails (the ideal $\hmin\to0$ at $t_0$). The \emph{self-similar?} lane
asks whether rescaled profiles collapse onto a single shape, as they do only inside the
intermediate similarity window. The \emph{universal?} lane asks what the shrinking region
forgets and what it retains, so that the local solution can shed the outer details yet still
carry a prefactor or a weakly decaying mode through the crossover. The \emph{regularised?}
lane asks which new physics arrests the ideal divergence at the smallest scale.}
\label{fig:toolkit}
\end{figure}

The most common mathematical signature of a dynamic interfacial singularity is a length
that tends to zero in a finite time. If $\hmin(t)$ is the minimum radius of a neck, we can write
\begin{equation}
\hmin(t) \sim A \ttc^\alpha \equiv A \tau^\alpha,
\label{eq:power-law}
\end{equation}
where $t_0$ is the time at singularity, $A$ is a prefactor, and $\alpha$ is an
exponent. We stress that though useful, a power law alone does not prove that the flow is universal.
The exponent may drift and require logarithmic corrections that multiply the leading power law by a slowly varying factor such as a logarithm of the scale ratio. In such cases, an exponent fitted over a finite window will not be a constant. Moreover, the prefactor may carry memory of the initial condition and different physical balances may be active in neighbouring time windows.

To compare the examples that follow, we will use the following four terms with precise and
distinct meanings (figure~\ref{fig:toolkit}).
\begin{enumerate}
\item A \emph{singularity} is the ideal failure of the smooth description, signalled by a
vanishing length, diverging curvature, unbounded stress, non-smooth field, or topological
obstruction.
\item A \emph{self-similar} solution is a special form in which profiles at different
times collapse after rescaling.
\item A \emph{universal} solution is one that attracts a class of initial and boundary
conditions, so that local behaviour no longer remembers most of the outer problem.
\item A \emph{regularisation} is the physical mechanism that stops the ideal divergence
by introducing a finite cutoff scale at which previously neglected physics enters.
\end{enumerate}
These ideas should not be conflated. For instance, diffusion provides self-similar solutions without any
finite-time singularity, while some singularities are not self-similar and some
self-similar solutions still retain a parameter fixed by history. Regularisation also does
not erase the singular dynamics, since an experiment can be cut off at small scales and
still display a large universal intermediate regime
\citep{EggersFontelos2015Singularities}.

For soft-matter examples, the distinctions above matter as singularities stage a
competition between forgetting and memory. The shrinking region can often isolate itself so
strongly that most details from outer scales become irrelevant.
For such problems, dimensional analysis is often the first predictive tool.
For a Newtonian liquid of density $\rho$, viscosity $\mu$, and surface tension $\gamma$, a
neck of radius $h$ has a capillary pressure scale $\gamma/h$. If inertia balances
capillarity, $\rho (\dot{h})^2 \sim \gamma/h$, which gives
\begin{equation}
\hmin \sim \left(\frac{\gamma}{\rho}\right)^{1/3} \ttc^{2/3}.
\label{eq:inertial-capillary}
\end{equation}
If viscosity balances capillarity, $\mu \dot{h}/h \sim \gamma/h$, which gives
\begin{equation}
\hmin \sim \frac{\gamma}{\mu}\ttc.
\label{eq:viscous-capillary}
\end{equation}
These estimates are deliberately incomplete as they ignore geometry, outer matching, and numerical prefactors. Yet they are useful in exposing a proposed dominant balance and predicting a power law that can be tested.
If the exponent is wrong, the assumed balance is wrong. On the other hand, if the exponent is right, the
remaining work is to determine where the law holds, what prefactor it carries, and how the
local solution matches the outer problem. A neck rarely stays in a single balance throughout the process. In interfacial problems the route to a singularity often passes through
crossovers between competing balances (section~\ref{sec:drop-pinch}).
Here, we note that some global constraints,
additional material timescales, or geometric asymmetries can still survive into the local
dynamics.
Drop breakup is a good example of this conceptual friction where the universal inviscid picture dominates (see section~\ref{sec:drop-pinch}), but experiments and computations also reveal memory effects in the final shape
and satellite-drop formation
\citep{Eggers1993Universal,Doshi2003Persistence}.

The second tool is matched asymptotics. A singular event is rarely described by one
scale. The \emph{outer} problem supplies forcing, geometry, and boundary conditions,
while the \emph{inner} problem contains the shrinking length on which the local balance
is made. Similarity is useful when this inner problem simplifies after rescaling;
universality is the stronger statement that the matching has forgotten most outer
details, and regularisation asks what still smaller scale replaces the
continuum limit.
For instance, in a moving contact line, the outer and inner length scales are provided by the droplet radius and the wedge scale near the contact line, respectively. In pinch-off, the analogous
outer scale may be a nozzle, bath, or initial bridge radius, but the inner scale is the
neck.

The third tool is tracking the evolving metric. For interfaces, this may be
a neck radius, a contact-line position, a curvature, or the topology of
the liquid domain. Indeed, a classical smooth-interface description cannot split or merge a
liquid body without passing through a singular event.
For ordered media, one instead tracks the distortion around a defect in a director
(the local preferred orientation) or phase field, down to the finite core that regularises the singularity.

Figure~\ref{fig:toolkit} collects these distinctions and tools into a working checklist. Faced
with a candidate singularity, we ask which length is shrinking, whether the
rescaled profiles collapse onto a single shape (self-similarity), what the shrinking region forgets (universality) and what
it retains, and which physical process cuts off the ideal divergence (regularisation).
A good singularity argument states which of these claims it is making and where that
claim is expected to hold. In a dripping faucet, as the thread thins, the four questions become
measurements of the neck radius, the profile shape, the memory of the outer drop, and the cutoff that
prevents a literal zero.

\section{Drop pinch-off as topology change}
\label{sec:drop-pinch}

\begin{figure}
\centering
\includegraphics[width=\textwidth]{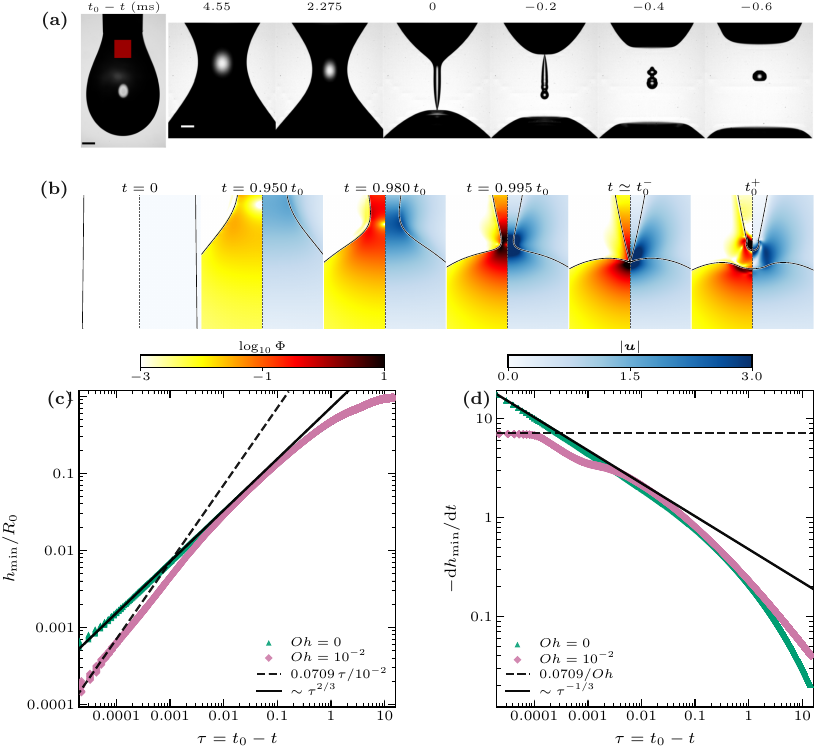}
\caption{Drop pinch-off, from experiment to thinning law.
(a)~A water drop pinching off from a nozzle, with the neck region (red box) enlarged in the
subsequent frames; scale bars $500\,\mu\mathrm{m}$ (black) and
$250\,\mu\mathrm{m}$ (white), and the labels give $\ttc$ in ms. The selected
frames have been reprinted with permission from
C.~I. Verschuur, A.~T. Oratis, V. Sanjay, and J.~H. Snoeijer, ``How elasticity affects
bubble pinch-off,'' \textit{Physical Review Fluids} \textbf{11}, 073302 (2026),
\href{https://doi.org/10.1103/5sp3-k5l2}{doi:10.1103/5sp3-k5l2}.
Copyright 2026 by the American Physical Society.
(b)~Newtonian axisymmetric simulation at $\Oh=10^{-2}$, each frame split into dissipation
$\log_{10}\Phi$ (left) and speed $|\boldsymbol{u}|$ (right)~\citep{dixit2025elasticpinchoffcode}.
(c)~Minimum neck radius $\hmin$ against time to pinch-off $\ttc$ for $\Oh=0$ and
$\Oh=10^{-2}$, with inertial ($\ttc^{2/3}$) and viscous ($\propto\ttc$) scalings.
(d)~The corresponding thinning rate $-\dot{h}_{\min}$ on log--log axes.}
\label{fig:drop-pinch}
\end{figure}

The breakup of a liquid thread is a canonical hydrodynamic singularity where surface
tension, trying to reduce interfacial area, amplifies the long-wave Rayleigh--Plateau
instability of a liquid cylinder once a neck has formed
\citep{Plateau1873Statique,Rayleigh1878Instability} (figure~\ref{fig:visual-dictionary}a). Capillary pressure then accelerates the local flow until
the neck radius reaches zero in a finite time; after that instant the
topology of the liquid domain has changed.

This event is central to soft matter for two reasons. It is easy to observe in a dripping
faucet (figure~\ref{fig:drop-pinch}a), an inkjet, a hanging (pendant) drop, or a collapsing capillary bridge, and it asks whether the neck
approaches a universal local shape independent of how the drop was made.

For a simple Newtonian liquid in air, the answer is yes, but often with varying
balances (figure~\ref{fig:drop-pinch}b). Which balance dominates at the outer scale $R$ is
set by the Ohnesorge number
\begin{equation}
\Oh=\frac{\mu}{\sqrt{\rho \gamma R}},
\label{eq:oh}
\end{equation}
named for Wolfgang von Ohnesorge, who introduced it to classify liquid-jet breakup,
$\Oh$ uses only fluid properties and size, not an imposed velocity
\citep{McKinleyRenardy2011WolfgangOhnesorge,Sanjay2022ViscousFreeSurfaceFlows}. It separates
inertial thinning at low $\Oh$ from viscous thinning at high $\Oh$.
The inviscid-capillary estimate in equation~\eqref{eq:inertial-capillary} gives
the familiar $2/3$ scaling, whereas a highly viscous thread approaches the
viscous-capillary scaling in equation~\eqref{eq:viscous-capillary} (figure~\ref{fig:drop-pinch}c). More refined theories show
that the neck profile is not determined by these exponents alone, because the axial length scale
matters, and so do imposed asymmetries and the matching to the outer flow; similarity
solutions can have selected shapes and constants that require solving the full local problem
\citep{DayHinchLister1998SelfSimilar,EggersFontelos2015Singularities}.

The central balance may change as the neck shrinks. A low-viscosity liquid can look
inertial over a large range of radii, then cross into a viscous regime once $\hmin$
becomes comparable to the viscous length
\begin{equation}
\ellnu=\frac{\mu^2}{\rho \gamma},
\label{eq:viscous-length}
\end{equation}
the scale below which viscosity becomes unavoidable in an otherwise inertial-capillary flow
(figure~\ref{fig:drop-pinch}c,d); a surrounding fluid can matter when its viscosity is not
negligible; finite gas density or pressure can modify the final stages; and satellite
drops form through secondary necks. No single universal law covers every experiment as the
neck usually passes through a sequence of local balances, each with its own range of
validity
\citep{CastrejonPita2015Plethora}.

The next question is what the neck remembers. If a neck region becomes sufficiently small and fast, one
expects it to decouple from the outer geometry. Yet experiments on two-fluid drop snap-off
and related breakup problems show that the local shape can retain memory of the initial
and boundary conditions longer than a naive universality argument would suggest
\citep{Doshi2003Persistence,Cohen_1999}. The shrinking region may forget many details,
while weakly decaying modes, imposed asymmetries, and global constraints survive into the
apparent similarity regime.

Pinch-off is powerful as a singularity experiment because the theory gives local
observables such as $\hmin(t)$, the profile shape, and the time to breakup, together with
the satellite drops formed afterwards.
Fitting the exponent is the easy part; the real tests come later on. Do the profiles
collapse onto a single shape, and does the exponent hold steady rather than drift? Does
the prefactor shift with nozzle size or outer fluid, and does the final satellite
distribution still remember the earlier neck shape?
A good singularity theory must predict these observables quantitatively.

\begin{figure}
\centering
\includegraphics[width=\textwidth]{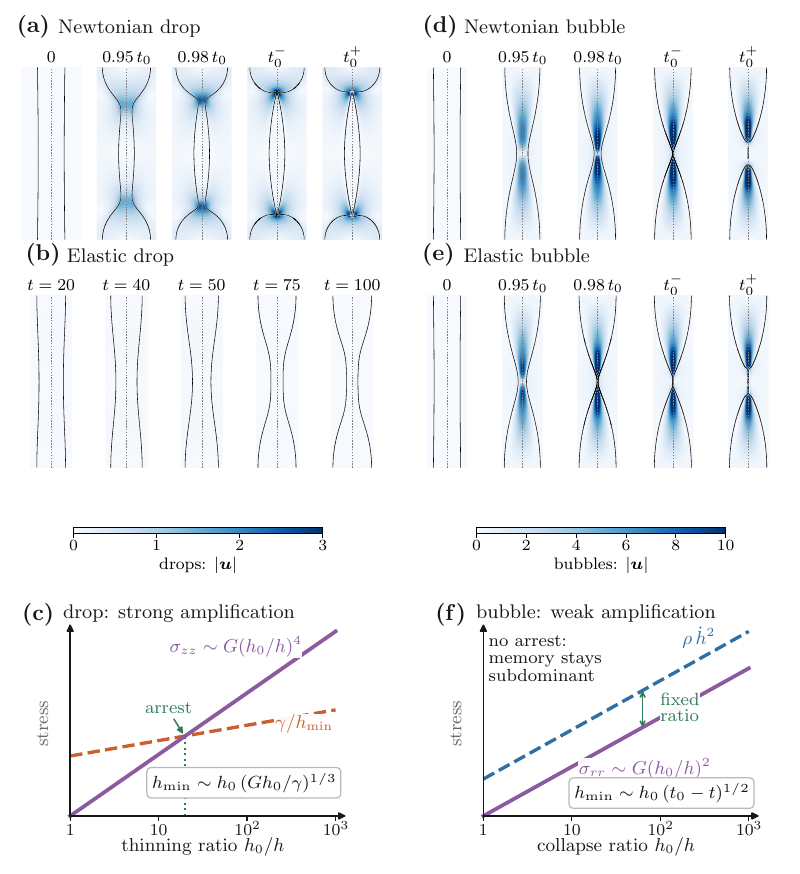}
\caption{Drop and bubble pinch-off with and without polymer elasticity. The left column
follows a drop (a--c) and the right a bubble (d--f). Snapshot panels show velocity magnitude
$|\boldsymbol{u}|$ in mirrored axisymmetric cross-sections of the full computational domain
\citep{dixit2025elasticpinchoffcode}; time advances from left to right, with colour scales
$0$--$3$ for drops and $0$--$10$ for bubbles. Panels (a,d) are Newtonian, whereas (b,e)
include dilute polymers: elasticity arrests the drop in a long-lived filament but leaves the
bubble close to its Newtonian collapse. Panels (c,f) compare the polymer stress with the
capillary or inertial driving; $G$ is the polymer modulus and $h_0/h$ the initial-to-current
neck-radius ratio. The scaling balances are discussed in the main text.}
\label{fig:drop-bubble}
\end{figure}

Complex fluids add another timescale to this event. A small amount of polymer can stretch
in the neck, generate elastic stress, and interrupt the Newtonian finite-time breakup
(figure~\ref{fig:drop-bubble}a,b). The singularity is then replaced by an elastocapillary
filament whose radius thins on the polymer relaxation time, because the axially stretched
polymers build a tensile stress that outgrows capillarity and arrests the thinning
(figure~\ref{fig:drop-bubble}c). Beads-on-a-string structures appear as the stretched thread and its stored elastic stress relax, leaving drops connected by thin filaments
\citep{AmaroucheneBonnMeunierKellay2001Inhibition,Eggers_2020_OldroydB}.

\section{Bubble versus drop pinch-off: same topology, different stress}
\label{sec:bubble-pinch}

At first glance, bubble pinch-off looks like drop pinch-off with the phases interchanged.
As a gas neck narrows in a surrounding liquid, the interface remains sharp, the neck radius
decreases, and topology changes.
That visual similarity is useful yet incomplete. A liquid thread pinches
because capillary pressure pulls liquid out of the neck, whereas
a bubble neck closes when the surrounding liquid inertia drives the radial
collapse of a cavity (figure~\ref{fig:drop-bubble}d). The singularity is interfacial in both
cases, but the dominant stress is different.

The simplest scaling for a collapsing bubble neck gives a radius close to
$\ttc^{1/2}$, but this square-root law is not a plain power law because logarithmic corrections enter through the long-ranged outer flow
\citep{GordilloSevillaRodriguezMartinezBazan2005BubblePinchOff,Gekle2009Approach}.
The exponent is therefore memorable, but the approach to universality is slow.
Initial geometry and perturbations can remain visible over experimentally accessible
ranges, so bubble pinch-off should not be reduced to a fitted slope.

The comparison with drop pinch-off is especially instructive once elasticity is added.
In a thinning liquid filament, polymers are stretched axially as the thread narrows, building a
tensile stress $\sigma_{zz}\sim G(h_0/h)^4$ that outgrows the capillary driving
and arrests the Newtonian route to breakup into an elastocapillary filament
(section~\ref{sec:drop-pinch}).
A collapsing bubble cavity gives the polymers far less influence.
Theory and simulation of dilute viscoelastic bubble pinch-off show that the radial
collapse stretches them only weakly, $\sigma_{rr}\sim G(h_0/h)^2$, so the elastic stress
grows no faster than the inertial driving $\rho\dot{h}^2$ and stays a fixed ratio below it
\citep{Verschuur2026ElasticityBubblePinchoff}.
The same polymer additive therefore arrests the drop but barely influences the bubble.
The elastic bubble collapses much as its Newtonian counterpart,
and the inertial law $\hmin\sim h_0\,(t_0-t)^{1/2}$ survives (figure~\ref{fig:drop-bubble}e,f).

Two pinching necks can have the same outline and different mechanics. In the drop,
capillarity drains an axially stretched liquid filament; in the bubble, the surrounding
liquid inertia collapses a cavity while material elements are stretched radially. The
relevant comparison is the flow field, the inertia-carrying phase, and the direction of
material stretching. Bubble pinch-off approaches universality only asymptotically, as is
common for hydrodynamic singularities.
The local neck may tend towards a universal form, but only through logarithmically slow
corrections.

\section{Coalescence as the reverse topology change}
\label{sec:coalescence}

\begin{figure}
\centering
\includegraphics[width=\textwidth]{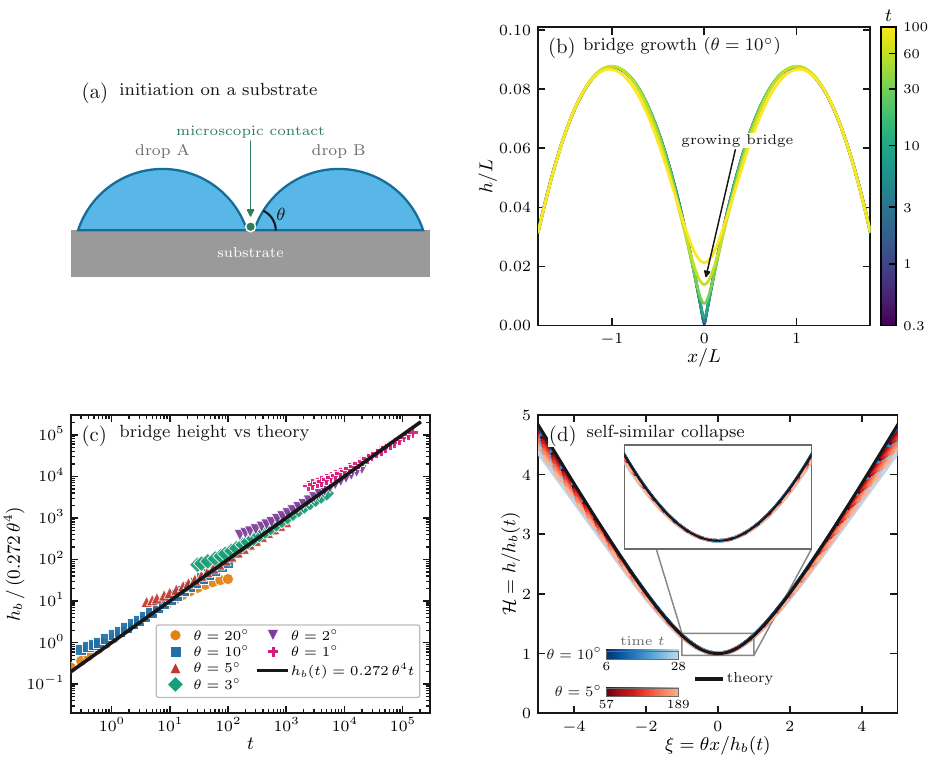}
\caption{Coalescence of two viscous \emph{sessile} (substrate-supported) drops. (a) The
drops first meet in a wedge set by the contact angle $\theta$, with microscopic physics
initiating the liquid bridge. (b) Clean-interface lubrication simulations for
$\theta=10^\circ$ show the bridge filling the wedge. (c) The bridge height follows
$h_b=0.272\,\theta^4t$; plotting $h_b/(0.272\,\theta^4)$ against $t$ collapses six
contact angles from
$1^\circ$ to $20^\circ$. (d) Profiles for $\theta=5^\circ$ and $10^\circ$, rescaled as
$\mathcal{H}=h/h_b$ and $\xi=\theta x/h_b$, collapse onto the black similarity solution;
the separate blue and red time keys run from early (dark) to late (light), and the inset
enlarges the neck. Simulations
are the surfactant-free limit of a lubrication model
\citep{talukdar2025coalescencecode}; the similarity law and coefficient are from
Hern\'{a}ndez-S\'{a}nchez et al.~\citep{hernandez2012symmetric}.}
\label{fig:coalescence}
\end{figure}

Pinch-off splits one fluid domain into two, whereas coalescence starts with two drops that
touch and become one through a bridge of radius $r(t)$ growing from a microscopic initial
size. The singularity now sits at the start rather than the end of the observable
motion.
The continuum description cannot describe the first molecular contact,
but once a bridge exists, capillarity drives rapid expansion
\citep{EggersListerStone1999Coalescence,EggersSprittlesSnoeijer2025CoalescenceDynamics}.

A common coalescence geometry is two sessile (substrate-supported) drops brought into contact.
Here, the geometry is fixed by the contact angle (figure~\ref{fig:coalescence}a). Two drops meet in a shallow wedge of half-angle
$\theta$. Because the bridge is much thinner than it is long, the flow is nearly parallel and
the governing equations simplify using the lubrication (thin-film) approximation. The bridge
then fills the wedge between the two caps (figure~\ref{fig:coalescence}b). Once the
microscopic initiation is forgotten, the problem keeps no length scale of its own, and the
growth is self-similar. The height grows linearly in time, $h_b(t) \simeq 0.272\,\theta^4 t$
in viscous-capillary units, and compensating by $0.272\theta^4$ collapses contact angles
from $1^\circ$ to $20^\circ$ onto a single line, each curve peeling away only once the bridge
feels the finite drop (figure~\ref{fig:coalescence}c). The local interface profile carries the same self-similar structure.
Rescaling vertical heights by $\theta^4 t$ and lateral distances by $\theta^3 t$ collapses bridge shapes
measured at different times and contact angles onto one master curve
(figure~\ref{fig:coalescence}d) \citep{hernandez2012symmetric}. By this stage, the
early bridge has forgotten both the microscopic initiation and the initial wedge angle.
A series of initial conditions is reduced to one local solution.
The complementary inertial, outer-fluid and
logarithmically corrected regimes, which dominate for low-viscosity drops coalescing in a
bulk, are reviewed in detail by Eggers, Sprittles \& Snoeijer
\citep{EggersSprittlesSnoeijer2025CoalescenceDynamics}; see also
\citep{Aarts_2005,Paulsen_2012}.

Coalescence can also turn the singular bridge into a local rheometer. Viscoelastic drops can
stretch material elements in the forming bridge and slow or reshape it through elastic
stresses \citep{Dekker_2022}; surfactants can make surface tension non-uniform, so that
Marangoni stresses oppose local interface creation or redistribution; and particles or
colloids trapped on the interface can give the bridge its own interfacial rheology. In
each case, the event asks whether the material can rearrange fast enough to keep up with
the new topology.

We emphasise that one should resist the urge to think of coalescence as simply pinch-off run backwards.
The equations of motion are time-reversal asymmetric because viscosity dissipates,
topology is initiated by microscopic contact rather than by a macroscopic neck reaching zero,
and the role of the surrounding gas or fluid can be different.
The conceptual link is topology change even though the mechanism is not identical.

\section{Moving contact lines: singularities that travel}
\label{sec:contact}

\begin{figure}
\centering
\includegraphics[width=\textwidth]{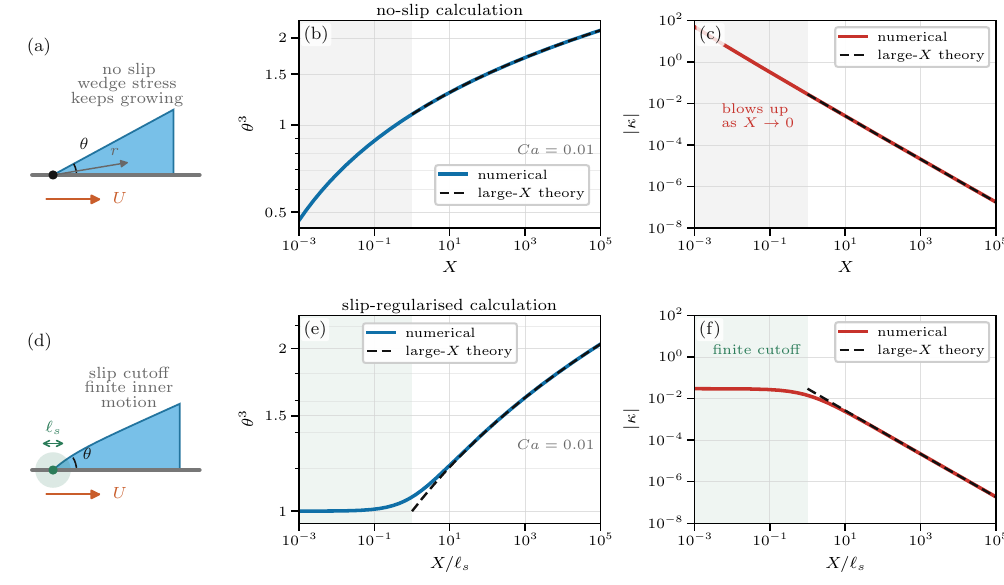}
\caption{A moving contact line as a spatial singularity. (a)~The no-slip wedge localises
stress near the moving line. (b,c)~Numerical integration of the no-slip contact-line
model~\citep{bhargava2026contactlinecode} over $10^{-3}<X<10^5$, where $X$ is the
distance from the contact line (made dimensionless) and $\Ca=\mu U/\gamma$ is the capillary
number, here $\Ca=0.01$. The apparent angle follows the large-$X$
Voinov prediction, while the curvature grows without bound as the contact line is
approached. (d)~A microscopic slip length cuts off the wedge. (e,f)~With that cutoff, the
angle and curvature saturate in the inner region but recover the same large-$X$ Voinov
behaviour outside the cutoff.}
\label{fig:contact-line}
\end{figure}

A contact line is the location where liquid, gas, and solid meet. At equilibrium this curve
is a geometrical boundary condition, but once it moves it becomes a hydrodynamic
singularity (figure~\ref{fig:visual-dictionary}b).
The classical no-slip model enforces that the fluid has the same velocity as the solid
boundary while the free surface is advected through the same point.
In the wedge near the contact line (figure~\ref{fig:contact-line}a), the shear stress
scales as $\mu U/r$, where $U$ is the contact-line speed and $r$ is the distance from the
line. The dissipation per unit length therefore becomes
\begin{equation}
\dot E \sim \int_{\ell}^{L} \mu U^2\,\frac{dr}{r}
        = \mu U^2 \ln\left(\frac{L}{\ell}\right).
\label{eq:contact-dissipation}
\end{equation}
If the continuum no-slip model were valid down to $r=0$, the force needed to move the
line would be logarithmically infinite. Taken literally, the contact line should not move.
That is the Huh--Scriven paradox
\citep{BonnEggersIndekeuMeunierRolley2009WettingSpreading,SnoeijerAndreotti2013MovingContactLines}.
Why, then, do raindrops slide down a window? Why do oil films retract into patches on a
cooking pan?

The same incompatibility appears in the lubrication description of a spreading drop. With
$X$ the distance from the contact line, $H(X)$ the local film height, and
$\Ca=\mu U/\gamma$ the capillary number, a reduced-order viscous bending model has the
form
\begin{equation}
\frac{d^3 H}{dX^3}+\frac{3\Ca}{H^2}=0,
\label{eq:contact-toy}
\end{equation}
where the first term is the curvature-gradient bending of the interface and the second is
the viscous resistance to
motion, which diverges as the film height $H\to0$. The equation tries to bend an interface
whose height vanishes at the line, and without a microscopic cutoff the slope and curvature
cannot be continued smoothly as $X\to0$ (figure~\ref{fig:contact-line}b,c). The
singularity is spatial rather than finite-time.

The resolution is to change the small-scale physics. A Navier slip length $\ell_s$
(figure~\ref{fig:contact-line}d) regularises the contact line by allowing fluid to move relative
to the wall. Other regularisations are also possible. A nanometric precursor film replaces
the sharp three-phase junction with a thin, continuously wetted layer, while a disjoining
pressure models intermolecular forces across that layer. Diffuse-interface and molecular-kinetic
descriptions instead replace the junction by a finite-width or
molecular-scale transition.
While the microscopic
closure is not unique, the macroscopic contact angle behaviour depends on it only through
the cutoff length.
In the viscous hydrodynamic regime, matching the inner regularised region to the outer drop yields the Cox--Voinov relation~\citep{SnoeijerAndreotti2013MovingContactLines}
\begin{equation}
\thetaa^3 = \thetae^3 + 9\Ca \ln\left(\frac{L}{\ell_m}\right),
\label{eq:cox-voinov}
\end{equation}
where $L$ is an outer length and $\ell_m$ is the microscopic cutoff. Equivalently,
$\theta^3(X)$ grows approximately linearly with
$\ln X$ away from the contact line. In the no-slip calculation shown in
figure~\ref{fig:contact-line}(b,c), with $\Ca=0.01$ and the far-field angle used as the unit
angle, the no-slip outer prediction can be written as
\begin{equation}
\theta^3 \simeq 1+9\Ca\ln(eX),
\qquad
\kappa \simeq \frac{d\theta}{dX}\simeq \frac{3\Ca}{X\theta^2}.
\label{eq:contact-voinov-curvature}
\end{equation}
The first expression is the Voinov logarithm in figure~\ref{fig:contact-line}(b), while the
second is the curvature blow-up in figure~\ref{fig:contact-line}(c).
Without a microscopic length, $\kappa\to\infty$ as $X\to0$. Introducing a slip length, or
any equivalent inner cutoff, replaces that unbounded small-$X$ limit by a finite inner
angle and a finite inner curvature (figure~\ref{fig:contact-line}e,f), while recovering
the same logarithmic law for
$X\gg\ell_s$. The logarithm is the important feature as the macroscopic angle depends on
microscopic physics, but only weakly, because each decade of length contributes and none
can be ignored completely. The memory is logarithmic, tied to the regularising scale
rather than to a surviving shape mode as in drop breakup.

Contact lines also introduce dynamical transitions. As a liquid film withdraws, its
receding edge (the dewetting front) can sharpen into a corner or leave narrow streams called
rivulets; advancing fronts can entrain air, while receding lines deposit films.
In some flows, two singular structures coexist, with a contact-line region and a
corner or cusp region that regularises the global geometry
\citep{PetersSnoeijerDaerrLimat2009Coexistence}. In such flows, the phrase ``the''
singularity becomes misleading because the continuum model contains several coupled
singular regions, not one isolated defect. As we have now encountered with other examples, regularisation can control measurable predictions.
If a wetting model hides the microscopic length, it may still predict trends correctly, but
it cannot give the full physical picture. Conversely, if a model focuses only on molecular motion and ignores the outer
hydrodynamic wedge, it misses why a macroscopic drop angle changes systematically with speed.

\section{Focusing singularities: cusps, tips, and films}
\label{sec:cusps}

\begin{figure}
\centering
\includegraphics[width=\textwidth]{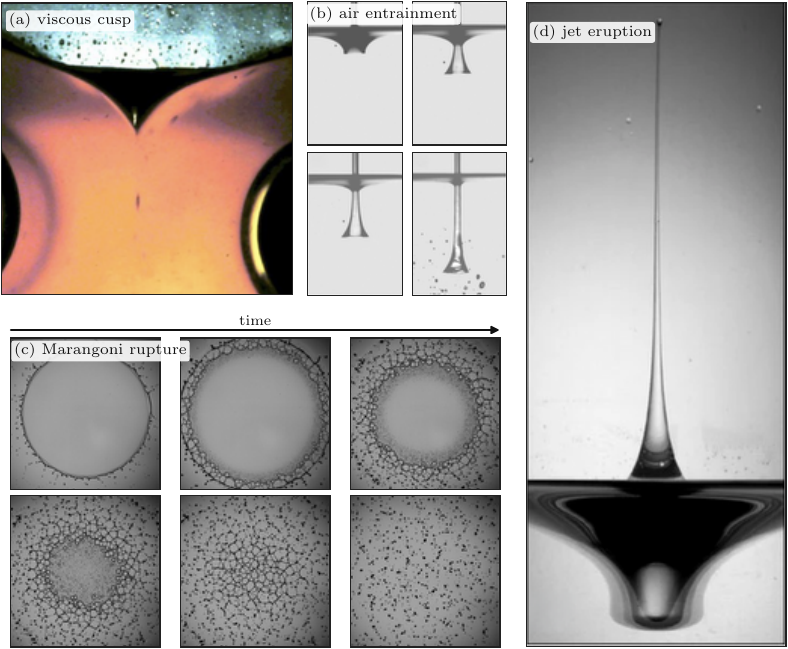}
\caption{Experimental views of interfacial localisation and output selection. (a) A
viscous roller flow pulls a free surface into a cusp. (b) A plunging viscous jet opens the
cusp into an entrained gas sheet. (c) Six high-speed frames show a $0.55\,\mathrm{mm}$
FC-84 drop striking a $340^{\circ}\mathrm{C}$ substrate at $3\,\mathrm{m\,s^{-1}}$; a
vapour layer supports the spreading sheet while thermocapillary stresses rupture it into
ligaments and droplets. (d) Inertial collapse of a surface-wave cavity focuses flow onto the
axis and launches a narrow jet; viscosity rounds its tip. Image credits are as
follows: Panel (a) has been reprinted with permission from
H.~K. Moffatt, ``Singularities in fluid mechanics,'' \textit{Physical Review Fluids}
\textbf{4}, 110502 (2019),
\href{https://doi.org/10.1103/PhysRevFluids.4.110502}{doi:10.1103/PhysRevFluids.4.110502}.
Copyright 2019 by the American Physical Society. The photographs in panel (b) have been
reprinted with permission from \'{E}. Lorenceau, D. Qu\'er\'e, and
J. Eggers, ``Air Entrainment by a Viscous Jet Plunging into a Bath,''
\textit{Physical Review Letters} \textbf{93}, 254501 (2004),
\href{https://doi.org/10.1103/PhysRevLett.93.254501}{doi:10.1103/PhysRevLett.93.254501}.
Copyright 2004 by the American Physical Society. Panel (c) is courtesy of Pierre
Chantelot, unpublished results, personal correspondence. Panel (d) is adapted
from B.~W. Zeff, B. Kleber, J. Fineberg, and D.~P. Lathrop, ``Singularity dynamics in
curvature collapse and jet eruption on a fluid surface,'' \textit{Nature} \textbf{403},
401--404 (2000), Springer Nature,
\href{https://doi.org/10.1038/35000151}{doi:10.1038/35000151}; reproduced with permission
from Springer Nature.}
\label{fig:focusing-output}
\end{figure}

Not every interfacial singularity changes topology. A free surface can sharpen into a cusp and
a thin film can rupture where its thickness is driven towards zero. The vanishing
length is then a radius of curvature or a local film thickness, respectively.
The question becomes how strongly the flow can localise curvature,
stress, or thickness.

Free-surface cusps at low Reynolds number give the cleanest geometry. Viscous flow pulls the
interface into a sharply curved tip (figure~\ref{fig:visual-dictionary}c; experiment in
figure~\ref{fig:focusing-output}a), while capillarity, finite interface thickness, or other
microscopic effects round the ideal cusp near its end
\citep{Jeong_1992,kumar2017bending}. The same competition between imposed strain and capillarity
sharpens the tips of selective withdrawal, where a sink flow draws an interface towards a point
\citep{CourrechDuPont_2006,CourrechDuPont_2020}. If the focusing is increased further,
a plunging jet or a fast contact line may drag a gas film into
the liquid, opening the cusp into entraining gas (figure~\ref{fig:focusing-output}b).
In coating, printing, or pouring, air entrainment may appear as a practical nuisance;
mechanically, it marks the point where a steady curved interface (the meniscus) can no longer keep both its apparent
angle and its curvature finite \citep{Eggers_2001_AirEntrainment,Lorenceau_2004}.

Thin-film rupture is the companion focusing singularity. Unlike the cusp, it is not tied to a single outer
balance. Macroscopic thinning may be driven by an interplay of capillarity, viscosity, and gravity,
and be further influenced by surfactants, or evaporation, while the final cutoff is microscopic, set by
van der Waals forces, disjoining pressure, or molecular discreteness \citep{Oron_1997}.
Rupture may then be deterministic, when an instability amplifies a thinning spot, or
stochastic, when thermal fluctuations matter close to the nanoscale
\citep{Kitavtsev_2018,Chatzigiannakis_2020}. The singularity is the organising limit; the
observed rupture is the regularised outcome.

The fragmentation sequence in figure~\ref{fig:focusing-output}(c) illustrates such a singularity.
A drop striking a substrate far above the liquid's boiling point rides a thin vapour
layer in the Leidenfrost regime and spreads into a sheet. On a hot enough surface, the sheet
does not simply recoil. Temperature gradients across the vapour layer drive thermocapillary
(Marangoni) stresses that thin the sheet locally, initiate rupture, and leave the rim and
interior to break into ligaments and then droplets. Here the shrinking length is the local
sheet thickness. The rupture location and cutoff are set by thermocapillary thinning,
thin-film forces, and the vapour layer that controls how thin the sheet becomes, rather than by
an outer flow pulling a passive interface into a cusp.

\section{Jets born from singular collapse}
\label{sec:jetting}

The singular events so far either change topology or grow from a singular contact. In many
flows, the same focusing has a second role as it converts a slow, distributed motion into a small
and fast output, such as a jet, a drop, a spray, or a short impulse on a wall. The question is
then how the outer flow feeds the singular region and what regularisation follows.

A collapsing cavity is a good prototype. When a surface depression caves in, inertia drives the
walls inward and focuses kinetic energy onto the axis, so velocity and surface curvature grow
together before a slender jet shoots out \citep{Zeff_2000}
(figure~\ref{fig:focusing-output}d). The collapse can become self-similar, but it does not
forget the outer problem completely. The cavity walls approach a cone whose semiangle $\beta$
is measured between one wall and the symmetry axis. Capillarity and viscosity select whether
the singular jetting regime is reached, the value of $\beta$, the inception time, and the final
cutoff. Once this conical cavity has formed, inertia dominates the local focusing until viscous--capillary stresses
regularise the smallest scale. Related conical focusing appears in cavitation, drop rebound, and
bubble bursting.

Bubble bursting is the obvious case as it clarifies why the jet is thinner and faster than
the naive inertial-capillary estimate $v_{\mathrm{jet}}\sim
\sqrt{\gamma/(\rho r_{\mathrm{jet}})}$, where $r_{\mathrm{jet}}$ is the jet radius.
The jet does not choose its
own scale. After the bubble ruptures, capillary waves run down the cavity and focus at its
base, sharpening the interface into a cone before the jet emerges; these waves, rather than
the surface tension of the jet itself, set the singular geometry
\citep{GordilloRodriguezRodriguez2019CapillaryWaves}. Their wavelength scales as $\Oh^{1/2}$,
so viscosity controls the cone they cut and makes the jet thinnest and fastest only in a narrow
window near $\Oh\approx0.03$, where the waves focus sharply before viscous damping removes them
\citep{GananCalvo_2017,Lai_2018,GordilloRodriguezRodriguez2019CapillaryWaves}.
Capillarity opens this window; viscosity closes it.

Within that window, the cone closes through a geometry-selected inertial similarity
\citep{GordilloRodriguezSanjay2026SelfSimilarWorthington}. Writing $\tau$ for the time
remaining until jet inception gives $r_{\mathrm{jet}}\sim\tau^{\alpha(\beta)}$ and
$v_{\mathrm{jet}}\sim\tau^{\alpha(\beta)-1}$, with $\beta$ selecting
$\alpha\simeq0.63$ near $\Oh=0.03$. Although close to the inertial--capillary value $2/3$,
this exponent implies a different force balance. Inertial--capillary scaling keeps the local
Weber number $\We=\rho v_{\mathrm{jet}}^2r_{\mathrm{jet}}/\gamma$ constant, whereas the
conical theory gives
$\We\sim\tau^{3\alpha-2}\sim r_{\mathrm{jet}}^{(3\alpha-2)/\alpha}\to\infty$ as
$r_{\mathrm{jet}}\to0$. Inertia therefore increasingly dominates the local focusing. The
rescaled interfaces collapse over more than two decades in dimensionless time, while
viscous--capillary stresses set the final cutoff at a few nanometres for micron-sized water
bubbles.
Polymer additives act on this process; once their relaxation is slower than jet formation,
they can suppress the jet drops or the jet itself, connecting the problem to the
elastocapillary response and local Deborah number discussed in section~\ref{sec:complex}
\citep{DixitOratisZinelisLohseSanjay2025ViscoelasticWorthington}.

Drop impact drives the same machinery from the opposite direction. The impact concentrates
pressure near the moving contact line and can launch a sheet or a jet
\citep{Cheng_2022_DropImpact}. On a non-wetting surface the force peaks again at jump-off, when
recoil refocuses the flow into a Worthington jet whose strength is fixed by inertia,
capillarity, and viscous damping
\citep{ZhangSanjayShiZhaoLvFengLohse2022ImpactForces,SanjayZhangLvLohse2025ViscosityImpactForces}.
Here the singular language is useful to understand the small length that sets the peak stress
and when a jet is ejected.

In inkjet printing, the pinch-off singularity selects the emitted drop volume and speed and
controls whether the residual ligament breaks into satellites \citep{Lohse_2022_Inkjet}. A
pressure pulse pushes the meniscus out of the nozzle and forms a liquid thread. The thread then
retracts and breaks. Surface tension drives the main breakup, but the same capillary retraction
can fragment the residual ligament and excite meniscus oscillations, so the waveform has to be
chosen for a particular nozzle and ink. Additives enter by changing the damping when viscosity
alone cannot keep the breakup on the single-drop branch.

A neck exponent is not yet an output prediction. It identifies a similarity regime, but it
does not say how much liquid leaves the singular region, whether as aerosol from a bursting
bubble or as the drop laid down by an inkjet. Impact teaches the same lesson in mechanical
form: the peak force depends on which small length focuses the stress.

\section{Material memory and additional cutoffs}
\label{sec:complex}

Elasticity, compared for drops and bubbles in section~\ref{sec:bubble-pinch}, is only the
first example of a broader complication. Near a singularity, polymer stretch, surface
concentration, particle contacts, or active stress can enter the dominant balance alongside
$\hmin$, curvature, pressure, or contact-line motion. A singular flow tests such variables
sharply because their evolution is driven on a shrinking length scale, where deformation
rates and gradients no longer remain outer-scale quantities.

For a polymer solution in a necking filament, the local measure is the Deborah
number, defined as the polymer relaxation time $\lambda$ divided by the local flow time.
The name recalls the prophetess Deborah's observation that ``the mountains flowed before
the Lord''. Over a sufficiently long observation time, even an apparently solid material can
flow \citep{Reiner1964DeborahNumber}.
Here, we note that the process time is changing during the
experiment, though the same product is often called a local Weissenberg number when the
emphasis is on deformation rate and polymer orientation
\citep{Poole2012DeborahWeissenberg}.
A useful estimate is
\begin{equation}
\De_\ell \sim \lambda \left|\frac{\dot{h}_{\min}}{\hmin}\right|.
\label{eq:de-local}
\end{equation}
Even if the imposed flow is initially weak, $\De_\ell$ can grow during pinch-off as the
necking rate accelerates, so material stretch that was passive at the outer scale becomes
part of the dominant balance. That crossover is the elastocapillary regularisation described in
section~\ref{sec:drop-pinch}. Once $\De_\ell$ reaches order unity, the stretched polymers carry the local stress, and the filament thins exponentially until finite extensibility or
solvent effects cut it off
\citep{AmaroucheneBonnMeunierKellay2001Inhibition,Eggers_2020_OldroydB,Deblais_2020}.

Memory can also enter through the creation and destruction of interface. A
surfactant-covered interface carries a concentration field set by adsorption and desorption
kinetics, surface diffusion, and sometimes surface viscosity or elasticity, and a singular
flow drives this field far from equilibrium in two opposing ways. Where fresh interface is
created faster than molecules can adsorb, as in a rapidly thinning neck, the new surface is
left locally clean and the dynamics revert towards the clean-interface law
\citep{McGoughBasaran2006RepeatedThreads}. Where the interface is compressed instead,
surfactant accumulates and the resulting surface-tension gradients drive Marangoni stresses
that resist the flow and slow the local thinning
\citep{KamatWagonerTheteBasaran2018Marangoni}. Particles can further influence this process. Once an adsorbed
monolayer is dense enough to jam, the interface buckles and crumples like an elastic skin
rather than relaxing as a simple surface
\citep{Subramaniam2005NonSpherical,Vella2004ParticleRaft}.

When the shrinking length reaches nanometres, singular dynamics are no longer insulated
from thermal fluctuations. A thin film may drain through a deterministic capillary-viscous
balance over most of its lifetime, but rupture when molecular forces and capillary
fluctuations produce a rare ``rogue'' nanowave \citep{sprittles2023rogue}. Coalescence
begins with a related problem. Before a bridge has a continuum radius, thermal motion
can bring two interfaces into molecular contact and nucleate the first liquid bridge
\citep{perumanath2019droplet}. The later hydrodynamics may then recover a deterministic
similarity law, in which case the fluctuation only chooses the time and place of the
event. If the near-singular state is weakly stable, that microscopic choice can remain
visible across the measured scaling window.

Active matter pushes the idea further by bringing in internal forcing instead of a passive memory.
For instance, in active nematics, defects do not merely ride on the flow they
generate; that flow can nucleate new defects or annihilate neighbours, while surviving
cores move under active forcing \citep{Giomi_2013,Shankar_2022}. The singular object
becomes a dynamical agent. Nonetheless, the singularities framework described here (see section~\ref{sec:toolkit}) can be used to describe
the divergence of continuum fields and how activity enhances or regularises them.

\section{Beyond interfaces}
\label{sec:beyond}

Interfacial flows have supplied the spine of the review because their singular regions can
be imaged directly. Beyond interfaces, the governing equations change, but the local
structure can remain, with a smooth description concentrating stress, curvature, topology, or
damage into a small region, and a material length prevents the ideal limit.

In a crumpled elastic sheet, the singular limit is set by the cost of stretching. For a
sheet of thickness $t$ and Young's modulus $E$, the bending modulus scales as
$B\sim Et^3$, whereas the stretching
modulus scales as $Y\sim Et$. Thin sheets therefore avoid in-plane strain over most of
their area and push the unavoidable stretching into ridges and vertices
(figure~\ref{fig:visual-dictionary}d). In the zero-thickness idealisation these folds and
points would sharpen without bound. Finite thickness cuts them off, sets the ridge width
and stress, and turns the focused region into a measurable object
\citep{Witten_2007,Cerda_1998}.

Soft solids pose the corresponding question with finite strain rather than surface
curvature. A compressed free surface can form creases or sulci, where self-contact
regularises the folded geometry, and a crack tip concentrates stress until a process zone
of damage and molecular rupture takes over
\citep{Hohlfeld_2011,Long_Hui_Gong_Bouchbinder_2021}.  In ordered soft materials,
topology can force an orientation or phase field to become undefined at a defect. Its finite
core regularises the divergent distortion \citep{Mermin_1979,Kamien_2002}. In active nematics,
active stresses move, create, and annihilate these defects
\citep{Giomi_2013,Shankar_2022}. Their governing fields and regularising physics therefore
differ from those of interfacial flows.

\begin{figure}
\centering
\includegraphics[width=\textwidth]{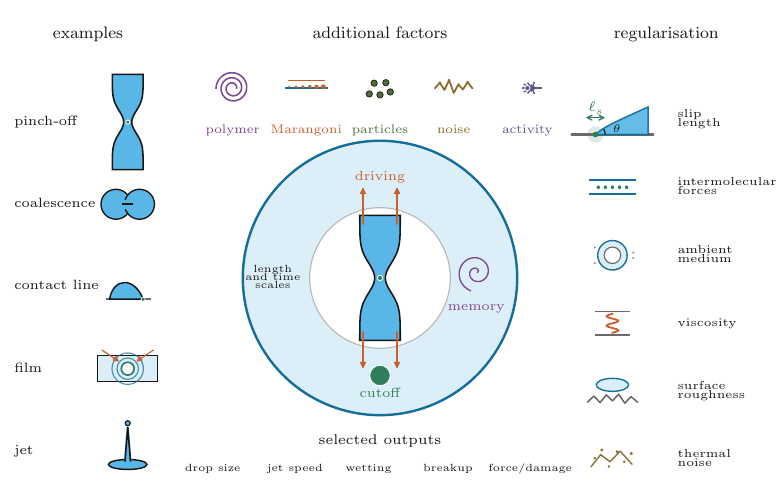}
\caption{Schematic of the singularity viewpoint developed in the review. Examples of
hydrodynamic singularities --- pinch-off, coalescence, contact lines, films, and jets --- feed a
local region where length and time scales, driving, material memory, and cutoff must be identified.
Additional factors such as polymers, Marangoni stresses, particles, fluctuations, and
activity can change the driving or the regularisation. Examples of regularising physics
include slip length, intermolecular forces, ambient medium, viscosity, surface roughness,
and thermal noise. The matched problem then selects measurable outputs including drop
size, jet speed, wetting transition, whether the structure breaks, and force or damage.}
\label{fig:broad-family}
\end{figure}

\section{Concluding remarks}
\label{sec:end}

A smooth continuum description usually fails locally rather than globally. The failure is
associated with a small region in which a length scale tends to zero, a curvature or stress
diverges, or an order parameter loses regularity. Identifying this region fixes the
questions that matter. One must determine the dominant balance, the shrinking length, and
the physical regularisation omitted by the outer model.

In this review, we have used interfacial flows to illustrate the different singularities in a continuum system.
Pinch-off shows most directly why an exponent is not enough. The scaling laws
$\hmin\propto\tau^{2/3}$ and $\hmin\propto\tau$ distinguish inertial--capillary and
viscous--capillary balances, yet the exponent fixes neither the crossover nor the prefactor,
and the satellite-drop volume can retain outer history. Singular Worthington jets
during bursting of bubbles give a more striking example. Although the predicted exponent $\alpha\simeq0.63$ lies close to the $2/3$ value from naive dimensional analysis, the local Weber number grows to infinity rather than remaining constant. A straight line on a
log--log plot can therefore conceal the wrong force balance. The moving
contact-line problem exemplifies a spatial singularity, where its apparent angle cannot be
predicted without connecting the outer wedge to a microscopic slip, molecular, or
diffuse-interface scale. At cusps and films, the final cutoff likewise dictates air entrainment and film rupture.

For soft materials, additional physics often enters before a nominal microscopic cutoff. In
a polymeric neck, $\De_\ell$ can grow as $\hmin$ falls; once it reaches order unity, elasticity
carries the local stress and changes the thinning law. Similarly, in a surfactant-laden flow, rapid interface creation leaves the surface locally clean, whereas compression builds Marangoni stresses.
Thermal fluctuations can further choose when and where a film ruptures or a bridge nucleates. In printing and coating, this is why satellite drops, air
entrainment, and film rupture depend on the liquid formulation, the gas, and the boundary
geometry as much as on the outer flow. Aerosols, mucus, cellular materials, soft robotics,
and stretchable solids pose the same issue in another form, with failure or output
selection initiated at a local stress concentration or thinning site rather than in the
bulk.

Figure~\ref{fig:broad-family} collects these ingredients. Outer forcing supplies mass,
momentum, stress, or topology to a region whose length scale is shrinking. Material memory
and additional physics can survive into this inner region. The cutoff fixes the inner
length, and the matched problem selects the measured output. This is why drop pinch-off,
contact-line motion, active defects, and elastic ridges can be compared without being
described by the same equations.

Elastic sheets, soft solids, foams, ordered media, and active systems sharpen this point
because the singular field is not always a free surface. The local variable may be
curvature, stress, orientation, density, or activity, and the cutoff may be set by
self-contact, process-zone physics, defect cores, grain-scale geometry, or fluctuations.
The useful question is what remains predictive once a continuum theory reaches its
limiting scale.

The remaining open problems are as much metrological as theoretical. For Newtonian
interfaces, many exponents are known. The more discriminating tests are shape collapse,
prefactor selection, stress or strain histories, and interactions between neighbouring
singular regions. For complex media, the question is whether memory, microstructure,
activity, or fluctuations only set the cutoff, or whether they enter the dominant balance
before the cutoff is reached. Simulations face the same constraint. A claim about
regularisation depends on resolution, constitutive closure, and the molecular or
mesoscopic physics left outside the continuum model.

The singularity viewpoint is therefore diagnostic. It identifies where a continuum theory
ceases to close on its own and which material physics must be added at that scale.

\section*{Acknowledgements}
The drop and bubble pinch-off simulations (figures~\ref{fig:drop-pinch} and~\ref{fig:drop-bubble})
use a code first written by Ayush Dixit, with later refinements to the adaptive mesh
refinement and to the configuration used here by Saumili Jana. The coalescence computations
(figure~\ref{fig:coalescence}) use a surfactant-lubrication code written with Jnandeep
Talukdar, and the moving-contact-line calculations (figure~\ref{fig:contact-line}) use codes
written with Aman Bhargava. I thank Pierre Chantelot for sharing the unpublished high-speed
images of \mbox{FC-84} drop fragmentation shown in figure~\ref{fig:focusing-output}(c). For
fruitful discussions I thank Aleksandr Bashkatov, Vincent Bertin, Aman Bhargava, Pierre
Chantelot, C.~Ricardo Constante-Amores, Ayush Dixit, Jie Feng, Jos\'{e} Manuel Gordillo,
Maziyar Jalaal, Saumili Jana, Dominik Krug, Detlef Lohse, Gareth McKinley, Alexandros Oratis,
Andrea Prosperetti, Javier Rodr\'{i}guez-Rodr\'{i}guez, Uddalok Sen, Jacco Snoeijer, Jnandeep
Talukdar, and Coen Verschuur. This work made use of the Hamilton HPC Service of Durham
University. The simulation work was also carried out on the national e-infrastructure of
SURFsara, a subsidiary of SURF cooperation, the collaborative ICT organisation for Dutch
education and research; this simulation work was sponsored by NWO~-- Domain Science for the
use of supercomputer facilities.

\section*{Disclosure statement}
No potential conflict of interest was reported by the author.

\section*{Funding}
The author acknowledges start-up funding from Durham University.

\section*{Data availability}
The figure-generation scripts, curated data used to rebuild the figures, and rendered
figure files are available in the public GitHub repository
\url{https://github.com/comphy-lab/Soft-Matter-Singularities-paper-figures}.

\bibliographystyle{tfnlm}
\bibliography{references}

@string{prl = "Phys. Rev. Lett."}

@string{jfm = "J. Fluid Mech."}

@string{rmp = "Rev. Mod. Phys."}

@string{arfm = "Annu. Rev. Fluid Mech."}

@string{pnas = "Proc. Natl. Acad. Sci. USA"}

@string{prfluids = "Phys. Rev. Fluids"}

@string{pof = "Phys. Fluids"}

@book{EggersFontelos2015Singularities,
  author = {Eggers, J. and Fontelos, M. A.},
  title     = {Singularities: Formation, Structure, and Propagation},
  series    = {Cambridge Texts in Applied Mathematics},
  publisher = {Cambridge University Press},
  address   = {Cambridge},
  year      = {2015},
  isbn      = {9781107098411},
  doi       = {10.1017/CBO9781316161692}
}

@article{Eggers1997Nonlinear,
  author = {Eggers, J.},
  title     = {Nonlinear dynamics and breakup of free-surface flows},
  journal = rmp,
  volume    = {69},
  number    = {3},
  pages     = {865--929},
  year      = {1997},
  month     = {jul},
  publisher = {American Physical Society},
  doi       = {10.1103/RevModPhys.69.865},
  url       = {https://link.aps.org/doi/10.1103/RevModPhys.69.865}
}

@article{Eggers1993Universal,
  author = {Eggers, J.},
  title     = {Universal Pinching of 3{D} Axisymmetric Free-Surface Flow},
  journal = prl,
  volume    = {71},
  number    = {21},
  pages     = {3458--3460},
  year      = {1993},
  month     = {nov},
  publisher = {American Physical Society},
  doi       = {10.1103/PhysRevLett.71.3458},
  url       = {https://link.aps.org/doi/10.1103/PhysRevLett.71.3458}
}

@article{DayHinchLister1998SelfSimilar,
  author = {Day, R. F. and Hinch, E. J. and Lister, J. R.},
  title     = {Self-Similar Capillary Pinchoff of an Inviscid Fluid},
  journal = prl,
  volume    = {80},
  number    = {4},
  pages     = {704--707},
  year      = {1998},
  month     = {jan},
  publisher = {American Physical Society},
  doi       = {10.1103/PhysRevLett.80.704},
  url       = {https://link.aps.org/doi/10.1103/PhysRevLett.80.704}
}

@article{Doshi2003Persistence,
  author = {Doshi, P. and Cohen, I. and Zhang, W. W. and Siegel, M. and Howell, P. and Basaran, O. A. and Nagel, S. R.},
  title     = {Persistence of Memory in Drop Breakup: The Breakdown of Universality},
  journal   = {Science},
  volume    = {302},
  number    = {5648},
  pages     = {1185--1188},
  year      = {2003},
  month     = {nov},
  publisher = {American Association for the Advancement of Science},
  doi       = {10.1126/science.1089272},
  url       = {https://www.science.org/doi/10.1126/science.1089272}
}

@article{EggersVillermaux2008PhysicsLiquidJets,
  author = {Eggers, J. and Villermaux, E.},
  title     = {Physics of Liquid Jets},
  journal = {Rep. Prog. Phys.},
  volume    = {71},
  number    = {3},
  pages     = {036601},
  year      = {2008},
  month     = {feb},
  publisher = {IOP Publishing},
  doi       = {10.1088/0034-4885/71/3/036601},
  url       = {https://doi.org/10.1088/0034-4885/71/3/036601}
}

@article{Cohen_1999,
  title = {Two Fluid Drop Snap-Off Problem: Experiments and Theory},
  author = {Cohen, I. and Brenner, M. P. and Eggers, J. and Nagel, S. R.},
  journal = prl,
  volume = {83},
  number = {6},
  pages = {1147--1150},
  year = {1999},
  month = aug,
  publisher = {American Physical Society},
  doi = {10.1103/PhysRevLett.83.1147},
  url = {https://doi.org/10.1103/PhysRevLett.83.1147}
}

@article{CastrejonPita2015Plethora,
  author = {Castrej{\'o}n-Pita, J. R. and Castrej{\'o}n-Pita, A. A. and Thete, S. S. and Sambath, K. and Hutchings, I. M. and Hinch, J. and Lister, J. R. and Basaran, O. A.},
  title = {Plethora of transitions during breakup of liquid filaments},
  journal = pnas,
  volume = {112},
  number = {15},
  pages = {4582--4587},
  year = {2015},
  month = {apr},
  publisher = {National Academy of Sciences},
  doi = {10.1073/pnas.1418541112},
  url = {https://www.pnas.org/doi/10.1073/pnas.1418541112}
}

@article{McKinleyRenardy2011WolfgangOhnesorge,
  author = {McKinley, Gareth H. and Renardy, Michael},
  title = {Wolfgang von {Ohnesorge}},
  journal = pof,
  volume = {23},
  number = {12},
  pages = {127101},
  year = {2011},
  month = dec,
  publisher = {AIP Publishing},
  doi = {10.1063/1.3663616},
  url = {https://doi.org/10.1063/1.3663616}
}

@phdthesis{Sanjay2022ViscousFreeSurfaceFlows,
  author = {Sanjay, Vatsal},
  title = {Viscous Free-Surface Flows},
  school = {University of Twente},
  address = {Enschede, The Netherlands},
  year = {2022},
  month = jul,
  isbn = {978-90-365-5407-7},
  doi = {10.3990/1.9789036554077},
  url = {https://doi.org/10.3990/1.9789036554077}
}

@article{GordilloSevillaRodriguezMartinezBazan2005BubblePinchOff,
  author = {Gordillo, J. M. and Sevilla, A. and Rodr{\'i}guez-Rodr{\'i}guez, J. and Mart{\'i}nez-Baz{\'a}n, C.},
  title = {Axisymmetric bubble pinch-off at high {Reynolds} numbers},
  journal = prl,
  volume = {95},
  number = {19},
  pages = {194501},
  year = {2005},
  month = {nov},
  publisher = {American Physical Society},
  doi = {10.1103/PhysRevLett.95.194501},
  url = {https://link.aps.org/doi/10.1103/PhysRevLett.95.194501}
}

@article{Gekle2009Approach,
  author = {Gekle, S. and Snoeijer, J. H. and Lohse, D. and van der Meer, D.},
  title     = {Approach to universality in axisymmetric bubble pinch-off},
  journal = {Phys. Rev. E},
  volume    = {80},
  number    = {3},
  pages     = {036305},
  year      = {2009},
  month     = {sep},
  publisher = {American Physical Society},
  doi       = {10.1103/PhysRevE.80.036305},
  url       = {https://link.aps.org/doi/10.1103/PhysRevE.80.036305}
}

@article{EggersListerStone1999Coalescence,
  author    = {Eggers, J. and Lister, J. R. and Stone, H. A.},
  title     = {Coalescence of liquid drops},
  journal = jfm,
  volume    = {401},
  pages     = {293--310},
  year      = {1999},
  month     = {dec},
  publisher = {Cambridge University Press},
  doi       = {10.1017/S002211209900662X},
  url       = {https://doi.org/10.1017/S002211209900662X}
}

@article{EggersSprittlesSnoeijer2025CoalescenceDynamics,
  author = {Eggers, J. and Sprittles, J. E. and Snoeijer, J. H.},
  title     = {Coalescence Dynamics},
  journal = arfm,
  volume    = {57},
  number    = {1},
  pages     = {61--87},
  year      = {2025},
  month     = {jan},
  publisher = {Annual Reviews},
  doi       = {10.1146/annurev-fluid-121021-044919},
  url       = {https://doi.org/10.1146/annurev-fluid-121021-044919}
}

@article{hernandez2012symmetric,
  title     = {Symmetric and asymmetric coalescence of drops on a substrate},
  author = {Hern{\'a}ndez-S{\'a}nchez, J. F. and Lubbers, L. A. and Eddi, A. and Snoeijer, J. H.},
  journal = prl,
  volume    = {109},
  number    = {18},
  pages     = {184502},
  year      = {2012},
  publisher = {American Physical Society},
  doi       = {10.1103/PhysRevLett.109.184502}
}

@article{Aarts_2005,
  title = {Hydrodynamics of Droplet Coalescence},
  author = {Aarts, D. G. A. L. and Lekkerkerker, H. N. W. and Guo, H. and Wegdam, G. H. and Bonn, D.},
  journal = prl,
  volume = {95},
  number = {16},
  pages = {164503},
  year = {2005},
  month = oct,
  publisher = {American Physical Society},
  doi = {10.1103/PhysRevLett.95.164503},
  url = {https://doi.org/10.1103/PhysRevLett.95.164503}
}

@article{Paulsen_2012,
  title = {The inexorable resistance of inertia determines the initial regime of drop coalescence},
  author = {Paulsen, J. D. and Burton, J. C. and Nagel, S. R. and Appathurai, S. and Harris, M. T. and Basaran, O. A.},
  journal = pnas,
  volume = {109},
  number = {18},
  pages = {6857--6861},
  year = {2012},
  month = apr,
  publisher = {Proceedings of the National Academy of Sciences},
  doi = {10.1073/pnas.1120775109},
  url = {https://doi.org/10.1073/pnas.1120775109}
}

@article{Dekker_2022,
  title = {When Elasticity Affects Drop Coalescence},
  author = {Dekker, P. J. and Hack, M. A. and Tewes, W. and Datt, C. and Bouillant, A. and Snoeijer, J. H.},
  journal = prl,
  volume = {128},
  number = {2},
  pages = {028004},
  year = {2022},
  month = jan,
  publisher = {American Physical Society},
  doi = {10.1103/PhysRevLett.128.028004},
  url = {https://doi.org/10.1103/PhysRevLett.128.028004}
}

@article{BonnEggersIndekeuMeunierRolley2009WettingSpreading,
  author = {Bonn, D. and Eggers, J. and Indekeu, J. and Meunier, J. and Rolley, E.},
  title     = {Wetting and Spreading},
  journal = rmp,
  volume    = {81},
  number    = {2},
  pages     = {739--805},
  year      = {2009},
  month     = {may},
  publisher = {American Physical Society},
  doi       = {10.1103/RevModPhys.81.739},
  url       = {https://doi.org/10.1103/RevModPhys.81.739}
}

@article{SnoeijerAndreotti2013MovingContactLines,
  author = {Snoeijer, J. H. and Andreotti, B.},
  title     = {Moving Contact Lines: Scales, Regimes, and Dynamical Transitions},
  journal = arfm,
  volume    = {45},
  pages     = {269--292},
  year      = {2013},
  doi       = {10.1146/annurev-fluid-011212-140734},
  url       = {https://doi.org/10.1146/annurev-fluid-011212-140734}
}

@article{PetersSnoeijerDaerrLimat2009Coexistence,
  author = {Peters, I. R. and Snoeijer, J. H. and Daerr, A. and Limat, L.},
  title     = {Coexistence of Two Singularities in Dewetting Flows: Regularizing the Corner Tip},
  journal = prl,
  volume    = {103},
  number    = {11},
  pages     = {114501},
  year      = {2009},
  month     = {sep},
  publisher = {American Physical Society},
  doi       = {10.1103/PhysRevLett.103.114501},
  url       = {https://link.aps.org/doi/10.1103/PhysRevLett.103.114501}
}

@article{Jeong_1992,
  title = {Free-surface cusps associated with flow at low Reynolds number},
  author = {Jeong, J.-T. and Moffatt, H. K.},
  journal = jfm,
  volume = {241},
  pages = {1--22},
  year = {1992},
  month = aug,
  publisher = {Cambridge University Press},
  doi = {10.1017/S0022112092001927},
  url = {https://doi.org/10.1017/S0022112092001927}
}

@article{Eggers_2001_AirEntrainment,
  title = {Air Entrainment through Free-Surface Cusps},
  author = {Eggers, J.},
  journal = prl,
  volume = {86},
  number = {19},
  pages = {4290--4293},
  year = {2001},
  month = may,
  publisher = {American Physical Society},
  doi = {10.1103/PhysRevLett.86.4290},
  url = {https://doi.org/10.1103/PhysRevLett.86.4290}
}

@article{Lorenceau_2004,
  title = {Air Entrainment by a Viscous Jet Plunging into a Bath},
  author = {Lorenceau, {\'E}. and Qu{\'e}r{\'e}, D. and Eggers, J.},
  journal = prl,
  volume = {93},
  number = {25},
  pages = {254501},
  year = {2004},
  month = dec,
  publisher = {American Physical Society},
  doi = {10.1103/PhysRevLett.93.254501},
  url = {https://doi.org/10.1103/PhysRevLett.93.254501}
}

@article{kumar2017bending,
  title = {Bending and growth of entrained air filament under converging and asymmetric rotational fields},
  author = {Kumar, P. and Das, A. K. and Mitra, S. K.},
  journal = pof,
  volume = {29},
  number = {2},
  pages = {022101},
  year = {2017},
  publisher = {AIP Publishing},
  doi = {10.1063/1.4975211},
  url = {https://doi.org/10.1063/1.4975211}
}

@article{CourrechDuPont_2006,
  title = {Sink Flow Deforms the Interface Between a Viscous Liquid and Air into a Tip Singularity},
  author = {Courrech du Pont, S. and Eggers, J.},
  journal = prl,
  volume = {96},
  number = {3},
  pages = {034501},
  year = {2006},
  month = jan,
  publisher = {American Physical Society},
  doi = {10.1103/PhysRevLett.96.034501},
  url = {https://doi.org/10.1103/PhysRevLett.96.034501}
}

@article{CourrechDuPont_2020,
  title = {Fluid interfaces with very sharp tips in viscous flow},
  author = {Courrech du Pont, S. and Eggers, J.},
  journal = pnas,
  volume = {117},
  number = {51},
  pages = {32238--32243},
  year = {2020},
  month = dec,
  publisher = {National Academy of Sciences},
  doi = {10.1073/pnas.2019287117},
  url = {https://doi.org/10.1073/pnas.2019287117}
}

@article{Oron_1997,
  title = {Long-scale evolution of thin liquid films},
  author = {Oron, A. and Davis, S. H. and Bankoff, S. G.},
  journal = rmp,
  volume = {69},
  number = {3},
  pages = {931--980},
  year = {1997},
  month = jul,
  publisher = {American Physical Society},
  doi = {10.1103/RevModPhys.69.931},
  url = {https://doi.org/10.1103/RevModPhys.69.931}
}

@article{Kitavtsev_2018,
  title = {Thermal rupture of a free liquid sheet},
  author = {Kitavtsev, G. and Fontelos, M. A. and Eggers, J.},
  journal = jfm,
  volume = {840},
  pages = {555--578},
  year = {2018},
  month = apr,
  publisher = {Cambridge University Press},
  doi = {10.1017/jfm.2018.74},
  url = {https://doi.org/10.1017/jfm.2018.74}
}

@article{Chatzigiannakis_2020,
  title = {Breakup of Thin Liquid Films: From Stochastic to Deterministic},
  author = {Chatzigiannakis, E. and Vermant, J.},
  journal = prl,
  volume = {125},
  number = {15},
  pages = {158001},
  year = {2020},
  month = oct,
  publisher = {American Physical Society},
  doi = {10.1103/PhysRevLett.125.158001},
  url = {https://doi.org/10.1103/PhysRevLett.125.158001}
}

@article{sprittles2023rogue,
  author = {Sprittles, J. E. and Liu, J. and Lockerby, D. A. and Grafke, T.},
  title = {Rogue nanowaves: A route to film rupture},
  journal = prfluids,
  volume = {8},
  number = {9},
  pages = {L092001},
  year = {2023},
  month = sep,
  publisher = {American Physical Society},
  doi = {10.1103/PhysRevFluids.8.L092001},
  url = {https://doi.org/10.1103/PhysRevFluids.8.L092001}
}

@article{perumanath2019droplet,
  author = {Perumanath, S. and Borg, M. K. and Chubynsky, M. V. and Sprittles, J. E. and Reese, J. M.},
  title = {Droplet coalescence is initiated by thermal motion},
  journal = prl,
  volume = {122},
  number = {10},
  pages = {104501},
  year = {2019},
  month = mar,
  publisher = {American Physical Society},
  doi = {10.1103/PhysRevLett.122.104501},
  url = {https://doi.org/10.1103/PhysRevLett.122.104501}
}

@article{Lai_2018,
  title = {Bubble Bursting: Universal Cavity and Jet Profiles},
  author = {Lai, C.-Y. and Eggers, J. and Deike, L.},
  journal = prl,
  volume = {121},
  number = {14},
  pages = {144501},
  year = {2018},
  month = oct,
  publisher = {American Physical Society},
  doi = {10.1103/PhysRevLett.121.144501},
  url = {https://doi.org/10.1103/PhysRevLett.121.144501}
}

@article{GananCalvo_2017,
  title = {Revision of Bubble Bursting: Universal Scaling Laws of Top Jet Drop Size and Speed},
  author = {Ganan-Calvo, A. M.},
  journal = prl,
  volume = {119},
  number = {20},
  pages = {204502},
  year = {2017},
  month = nov,
  publisher = {American Physical Society},
  doi = {10.1103/PhysRevLett.119.204502},
  url = {https://doi.org/10.1103/PhysRevLett.119.204502}
}

@article{Zeff_2000,
  title = {Singularity dynamics in curvature collapse and jet eruption on a fluid surface},
  author = {Zeff, B. W. and Kleber, B. and Fineberg, J. and Lathrop, D. P.},
  journal = {Nature},
  volume = {403},
  pages = {401--404},
  year = {2000},
  month = jan,
  publisher = {Springer Nature},
  doi = {10.1038/35000151},
  url = {https://doi.org/10.1038/35000151}
}

@article{Cheng_2022_DropImpact,
  title = {Drop Impact Dynamics: Impact Force and Stress Distributions},
  author = {Cheng, X. and Sun, T.-P. and Gordillo, L.},
  journal = arfm,
  volume = {54},
  number = {1},
  pages = {57--81},
  year = {2022},
  month = jan,
  publisher = {Annual Reviews},
  doi = {10.1146/annurev-fluid-030321-103941},
  url = {https://doi.org/10.1146/annurev-fluid-030321-103941}
}

@article{ZhangSanjayShiZhaoLvFengLohse2022ImpactForces,
  title = {Impact Forces of Water Drops Falling on Superhydrophobic Surfaces},
  author = {Zhang, B. and Sanjay, V. and Shi, S. and Zhao, Y. and Lv, C. and Feng, X.-Q. and Lohse, D.},
  journal = prl,
  volume = {129},
  number = {10},
  pages = {104501},
  year = {2022},
  month = aug,
  publisher = {American Physical Society},
  doi = {10.1103/PhysRevLett.129.104501},
  url = {https://link.aps.org/doi/10.1103/PhysRevLett.129.104501}
}

@article{SanjayZhangLvLohse2025ViscosityImpactForces,
  title = {The role of viscosity on drop impact forces on non-wetting surfaces},
  author = {Sanjay, V. and Zhang, B. and Lv, C. and Lohse, D.},
  journal = jfm,
  volume = {1004},
  pages = {A6},
  year = {2025},
  month = jan,
  publisher = {Cambridge University Press},
  doi = {10.1017/jfm.2024.982},
  url = {https://doi.org/10.1017/jfm.2024.982}
}

@article{Lohse_2022_Inkjet,
  title = {Fundamental Fluid Dynamics Challenges in Inkjet Printing},
  author = {Lohse, D.},
  journal = arfm,
  volume = {54},
  number = {1},
  pages = {349--382},
  year = {2022},
  month = jan,
  publisher = {Annual Reviews},
  doi = {10.1146/annurev-fluid-022321-114001},
  url = {https://doi.org/10.1146/annurev-fluid-022321-114001}
}

@misc{GordilloRodriguezSanjay2026SelfSimilarWorthington,
  author = {Gordillo, Jos{\'e} M. and Rodr{\'i}guez-Rodr{\'i}guez, Javier and Sanjay, Vatsal},
  title = {Self-similar {Worthington} jets},
  year = {2026},
  month = jul,
  eprint = {2607.08972},
  archivePrefix = {arXiv},
  primaryClass = {physics.flu-dyn},
  doi = {10.48550/arXiv.2607.08972},
  url = {https://arxiv.org/abs/2607.08972}
}

@article{GordilloRodriguezRodriguez2019CapillaryWaves,
  title = {Capillary waves control the ejection of bubble bursting jets},
  author = {Gordillo, J. M. and Rodr{\'i}guez-Rodr{\'i}guez, J.},
  journal = jfm,
  volume = {867},
  pages = {556--571},
  year = {2019},
  month = may,
  publisher = {Cambridge University Press},
  doi = {10.1017/jfm.2019.161},
  url = {https://doi.org/10.1017/jfm.2019.161}
}

@article{DixitOratisZinelisLohseSanjay2025ViscoelasticWorthington,
  title = {Viscoelastic {Worthington} jets and droplets produced by bursting bubbles},
  author = {Dixit, A. and Oratis, A. and Zinelis, K. and Lohse, D. and Sanjay, V.},
  journal = jfm,
  volume = {1010},
  pages = {A2},
  year = {2025},
  month = may,
  publisher = {Cambridge University Press},
  doi = {10.1017/jfm.2025.237},
  url = {https://doi.org/10.1017/jfm.2025.237}
}

@article{AmaroucheneBonnMeunierKellay2001Inhibition,
  author    = {Amarouchene, Y. and Bonn, D. and Meunier, J. and Kellay, H.},
  title     = {Inhibition of the Finite-Time Singularity during Droplet Fission of a Polymeric Fluid},
  journal = prl,
  volume    = {86},
  number    = {16},
  pages     = {3558--3561},
  year      = {2001},
  month     = {apr},
  publisher = {American Physical Society},
  doi       = {10.1103/PhysRevLett.86.3558},
  url       = {https://link.aps.org/doi/10.1103/PhysRevLett.86.3558}
}

@article{Eggers_2020_OldroydB,
  title = {Self-similar breakup of polymeric threads as described by the Oldroyd-B model},
  author = {Eggers, J. and Herrada, M. A. and Snoeijer, J. H.},
  journal = jfm,
  volume = {887},
  pages = {A19},
  year = {2020},
  publisher = {Cambridge University Press},
  doi = {10.1017/jfm.2020.12},
  url = {https://doi.org/10.1017/jfm.2020.12}
}

@article{Deblais_2020,
  title = {Self-similarity in the breakup of very dilute viscoelastic solutions},
  author = {Deblais, A. and Herrada, M. A. and Eggers, J. and Bonn, D.},
  journal = jfm,
  volume = {904},
  pages = {R2},
  year = {2020},
  month = oct,
  publisher = {Cambridge University Press},
  doi = {10.1017/jfm.2020.765},
  url = {https://doi.org/10.1017/jfm.2020.765}
}

@article{Poole2012DeborahWeissenberg,
  author = {Poole, R. J.},
  title = {The {Deborah} and {Weissenberg} numbers},
  journal = {Rheol. Bull.},
  volume = {53},
  number = {2},
  pages = {32--39},
  year = {2012},
  url = {https://pcwww.liv.ac.uk/~robpoole/PAPERS/POOLE\_45.pdf}
}

@article{Reiner1964DeborahNumber,
  author = {Reiner, M.},
  title = {The {Deborah} Number},
  journal = {Phys. Today},
  volume = {17},
  number = {1},
  pages = {62},
  year = {1964},
  month = jan,
  publisher = {AIP Publishing},
  doi = {10.1063/1.3051374},
  url = {https://doi.org/10.1063/1.3051374}
}

@article{Verschuur2026ElasticityBubblePinchoff,
  author = {Verschuur, C. I. and Oratis, A. T. and Sanjay, V. and Snoeijer, J. H.},
  title = {How elasticity affects bubble pinch-off},
  journal = prfluids,
  volume = {11},
  number = {7},
  pages = {073302},
  year = {2026},
  month = jul,
  publisher = {American Physical Society},
  doi = {10.1103/5sp3-k5l2},
  url = {https://doi.org/10.1103/5sp3-k5l2}
}

@article{Witten_2007,
  title = {Stress focusing in elastic sheets},
  author = {Witten, T. A.},
  journal = rmp,
  volume = {79},
  number = {2},
  pages = {643--675},
  year = {2007},
  month = apr,
  publisher = {American Physical Society},
  doi = {10.1103/RevModPhys.79.643},
  url = {https://doi.org/10.1103/RevModPhys.79.643}
}

@article{Cerda_1998,
  title = {Conical Surfaces and Crescent Singularities in Crumpled Sheets},
  author = {Cerda, E. and Mahadevan, L.},
  journal = prl,
  volume = {80},
  number = {11},
  pages = {2358--2361},
  year = {1998},
  month = mar,
  publisher = {American Physical Society (APS)},
  doi = {10.1103/PhysRevLett.80.2358},
  url = {https://doi.org/10.1103/PhysRevLett.80.2358}
}

@article{Hohlfeld_2011,
  title = {Unfolding the Sulcus},
  author = {Hohlfeld, E. and Mahadevan, L.},
  journal = prl,
  volume = {106},
  number = {10},
  pages = {105702},
  year = {2011},
  month = mar,
  publisher = {American Physical Society (APS)},
  doi = {10.1103/PhysRevLett.106.105702},
  url = {https://doi.org/10.1103/PhysRevLett.106.105702}
}

@article{Long_Hui_Gong_Bouchbinder_2021,
  title = {The Fracture of Highly Deformable Soft Materials: A Tale of Two Length Scales},
  author = {Long, R. and Hui, C.-Y. and Gong, J. P. and Bouchbinder, E.},
  journal = {Annu. Rev. Condens. Matter Phys.},
  volume = {12},
  number = {1},
  pages = {71--94},
  year = {2021},
  month = mar,
  publisher = {Annual Reviews},
  doi = {10.1146/annurev-conmatphys-042020-023937},
  url = {https://doi.org/10.1146/annurev-conmatphys-042020-023937},
  eprint = {2004.03159},
  archivePrefix = {arXiv},
  primaryClass = {cond-mat.soft}
}

@article{Mermin_1979,
  title = {The topological theory of defects in ordered media},
  author = {Mermin, N. D.},
  journal = rmp,
  volume = {51},
  number = {3},
  pages = {591--648},
  year = {1979},
  month = jul,
  publisher = {American Physical Society},
  doi = {10.1103/RevModPhys.51.591},
  url = {https://doi.org/10.1103/RevModPhys.51.591}
}

@article{Kamien_2002,
  title = {The geometry of soft materials: a primer},
  author = {Kamien, R. D.},
  journal = rmp,
  volume = {74},
  number = {4},
  pages = {953--971},
  year = {2002},
  month = oct,
  publisher = {American Physical Society},
  doi = {10.1103/RevModPhys.74.953},
  eprint = {cond-mat/0203127},
  archivePrefix = {arXiv},
  url = {https://doi.org/10.1103/RevModPhys.74.953}
}

@article{Shankar_2022,
  title = {Topological active matter},
  author = {Shankar, S. and Souslov, A. and Bowick, M. J. and Marchetti, M. C. and Vitelli, V.},
  journal = {Nature Rev. Phys.},
  volume = {4},
  number = {6},
  pages = {380--398},
  year = {2022},
  month = may,
  publisher = {Springer Science and Business Media LLC},
  doi = {10.1038/s42254-022-00445-3},
  eprint = {2010.00364},
  archivePrefix = {arXiv},
  url = {https://doi.org/10.1038/s42254-022-00445-3}
}

@article{Giomi_2013,
  title = {Defect Annihilation and Proliferation in Active Nematics},
  author = {Giomi, L. and Bowick, M. J. and Ma, X. and Marchetti, M. C.},
  journal = prl,
  volume = {110},
  number = {22},
  pages = {228101},
  year = {2013},
  month = may,
  publisher = {American Physical Society (APS)},
  doi = {10.1103/PhysRevLett.110.228101},
  eprint = {1303.4720},
  archivePrefix = {arXiv},
  url = {https://doi.org/10.1103/PhysRevLett.110.228101}
}

@misc{dixit2025elasticpinchoffcode,
  author       = {Sanjay, V. and collaborators},
  title        = {Code repository: {Elastic pinch-off}},
  year         = {2025},
  howpublished = {\url{https://github.com/comphy-lab/ElasticPinchOff}}
}

@misc{talukdar2025coalescencecode,
  author       = {Talukdar, J. and Rocha, D. and Sanjay, V.},
  title        = {Code repository: {Coalescence with surfactants}},
  year         = {2025},
  doi          = {10.5281/zenodo.18042315},
  howpublished = {\url{https://github.com/comphy-lab/coalescence-with-surfactants}}
}

@misc{bhargava2026contactlinecode,
  author       = {Bhargava, A. and Sanjay, V.},
  title        = {Code repository: {Contact-Line-101}},
  year         = {2026},
  howpublished = {\url{https://github.com/comphy-lab/Contact-Line-101}}
}

@article{McGoughBasaran2006RepeatedThreads,
  author    = {McGough, P. T. and Basaran, O. A.},
  title     = {Repeated Formation of Fluid Threads in Breakup of a Surfactant-Covered Jet},
  journal = prl,
  volume    = {96},
  number    = {5},
  pages     = {054502},
  year      = {2006},
  month     = {feb},
  publisher = {American Physical Society},
  doi       = {10.1103/PhysRevLett.96.054502},
  url       = {https://doi.org/10.1103/PhysRevLett.96.054502}
}

@article{KamatWagonerTheteBasaran2018Marangoni,
  author    = {Kamat, P. M. and Wagoner, B. W. and Thete, S. S. and Basaran, O. A.},
  title     = {Role of {Marangoni} stress during breakup of surfactant-covered liquid threads: Reduced rates of thinning and microthread cascades},
  journal = prfluids,
  volume    = {3},
  number    = {4},
  pages     = {043602},
  year      = {2018},
  month     = {apr},
  publisher = {American Physical Society},
  doi       = {10.1103/PhysRevFluids.3.043602},
  url       = {https://doi.org/10.1103/PhysRevFluids.3.043602}
}

@article{Subramaniam2005NonSpherical,
  author    = {Subramaniam, A. B. and Abkarian, M. and Mahadevan, L. and Stone, H. A.},
  title     = {Non-spherical bubbles},
  journal   = {Nature},
  volume    = {438},
  number    = {7070},
  pages     = {930},
  year      = {2005},
  month     = {dec},
  publisher = {Nature Publishing Group},
  doi       = {10.1038/438930a},
  url       = {https://doi.org/10.1038/438930a}
}

@article{Vella2004ParticleRaft,
  author    = {Vella, D. and Aussillous, P. and Mahadevan, L.},
  title     = {Elasticity of an interfacial particle raft},
  journal   = {Europhys. Lett.},
  volume    = {68},
  number    = {2},
  pages     = {212--218},
  year      = {2004},
  month     = {oct},
  publisher = {IOP Publishing},
  doi       = {10.1209/epl/i2004-10202-x},
  url       = {https://doi.org/10.1209/epl/i2004-10202-x}
}

@book{Plateau1873Statique,
  author    = {Plateau, J. A. F.},
  title     = {Statique exp\'{e}rimentale et th\'{e}orique des liquides soumis aux seules forces mol\'{e}culaires},
  publisher = {Gauthier-Villars},
  address   = {Paris},
  year      = {1873}
}

@article{Rayleigh1878Instability,
  author    = {{Lord Rayleigh}},
  title     = {On the instability of jets},
  journal   = {Proc. London Math. Soc.},
  volume    = {s1-10},
  number    = {1},
  pages     = {4--13},
  year      = {1878},
  doi       = {10.1112/plms/s1-10.1.4},
  url       = {https://doi.org/10.1112/plms/s1-10.1.4}
}

\end{document}